\documentclass[traditabstract]{aa} 

\usepackage[dvips]{graphicx}          
\usepackage[comma,authoryear]{natbib} 
\bibpunct{(}{)}{,}{a}{}{,}            

\begin{document}

   \title{Seismic signature of helium ionization in the Sun and stars }


   \author{G. Houdek\inst{1}
              \and
           D.O. Gough\inst{2,3}
          }

   \institute{Institute of Astronomy, University of Vienna,
              A-1180 Vienna, Austria,
              \email{guenter.houdek@univie.ac.at}
              \and
              Institute of Astronomy, University of Cambridge,
              Cambridge CB3 0HA, UK,
              \email{douglas@ast.cam.ac.uk}
              \and
              Department of Applied Mathematics and Theoretical Physics, 
              University of Cambridge CB3\,0WA, UK 
             }

  \abstract
{We present a summary of an asteroseismic signature of helium ionization
 reported by Houdek \& Gough (2007, 2008, 2009) for
 low-degree p modes in solar-type stars, and illustrate 
 its applications for asteroseismic diagnoses.}


   \maketitle

\section{Introduction}

Abrupt variation in the stratification of a star (relative to the 
scale of the inverse radial wavenumber of a seismic mode of oscillation), 
such as that resulting from the (smooth, albeit acoustically relatively abrupt) 
depression in the first adiabatic exponent 
$\gamma_1=(\partial {\ln p}/\partial{\ln\rho})_s$ caused by the ionization 
of helium, where $p$, $\rho$ and $s$ are pressure, density and specific 
entropy, or from the sharp transition from radiative to convective heat 
transport at the base of the convection zone, induces small-amplitude 
oscillatory components (with respect to frequency) in the spacing of the 
cyclic eigenfrequencies $\nu_i$ of seismic modes, with $i=(n,l)$, where $n$ 
and $l$ are order and degree respectively.
We shall call such abrupt variations in the stellar stratification
an acoustic glitch. Our interest is principally in the near-surface glitch 
caused by helium ionization. 

\section{The seismic diagnostic and discussion}
It was demonstrated by \citet{Houdek07} that the 
variation of the sound speed induced by helium ionization enables one to 
determine from the low-degree eigenfrequencies a measure that is almost 
proportional to the helium abundance, with little contamination from other 
properties of the structure of the star. 
The deviation
\begin{equation}
\delta\nu_i:=\nu_i-\nu_{{\rm s}i}
\end{equation}
of the eigenfrequency from the
corresponding frequency $\nu_{{\rm s}i}$ of a similar smoothly stratified, i.e.
glitch-free, star is the indicator of a measure of the helium abundance $Y$. 
The individual frequency contributions arising 
from the acoustic glitches to the total frequency contribution $\delta\nu_i$ are:
\begin{equation}
\delta\nu_i=\delta_\gamma\nu_i+\delta_{\rm c}\nu_i+\delta_{\rm u}\nu_i\,,
\label{e:all_glitches}
\end{equation}
\begin{figure}[tbp]
     \begin{center}
       \includegraphics[width=7.8cm]{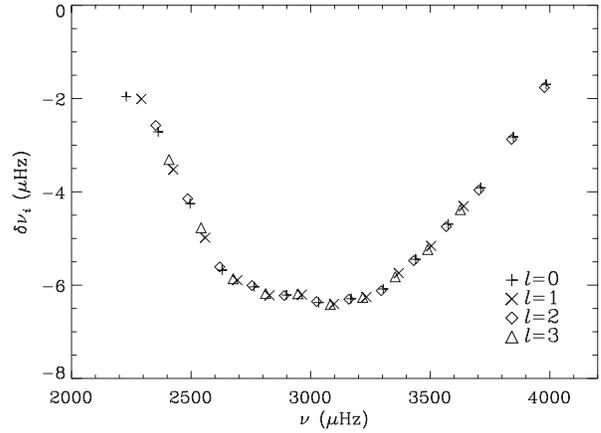}
     \end{center}
     \caption{The symbols denote contributions $\delta\nu_i$ to the frequencies
              $\nu_i$ produced by the acoustic glitches of the Sun.}
     \label{fig:one_column}
   \end{figure}
where $\delta_\gamma\nu_i$ and $\delta_{\rm c}\nu_i$ arise respectively from 
the variation in $\gamma_1$ induced by helium ionization and from the acoustic 
glitch at the base of the convection zone caused by a near discontinuity in 
the second derivative of density. Approximate expressions for 
$\delta_\gamma\nu_i$ and $\delta_{\rm c}\nu_i$ were presented by 
\citet{Houdek07, Houdek08}. The additional upper-glitch component
$\delta_{\rm u}\nu_i$, which is produced by the ionization of hydrogen and the 
upper superadiabatic boundary layer of the envelope convection zone, is 
difficult to model. \citet{Houdek07, Houdek08} estimated all three glitch 
components by fitting to second differences 
$\Delta_{2i}\nu:=\nu_{n-1,l}-2\nu_{n,l}+\nu_{n+1,l}$
of the observed frequencies $\nu_i$ the seismic diagnostic
\begin{equation}
\Delta_{2i}\nu
\simeq\Delta_{2i}(\delta_\gamma\nu+\delta_{\rm c}\nu+\delta_{\rm u}\nu)\,,
\end{equation}
and represented the upper-glitch component by a series of inverse powers
of $\nu_i$, i.e. $\Delta_{2i}\delta_{\rm u}\nu=\sum_{k=0}^3a_k\nu_i^{-k}$,
from which the four coefficients $a_k$, additional to seven more
coefficients in $\delta_\gamma\nu_i$ and $\delta_{\rm c}\nu_i$, were determined.
The upper glitch frequency component is represented by \citep{Houdek09}

   \begin{figure*}[htbp]
     \begin{center}
       \includegraphics[width=18.0cm]{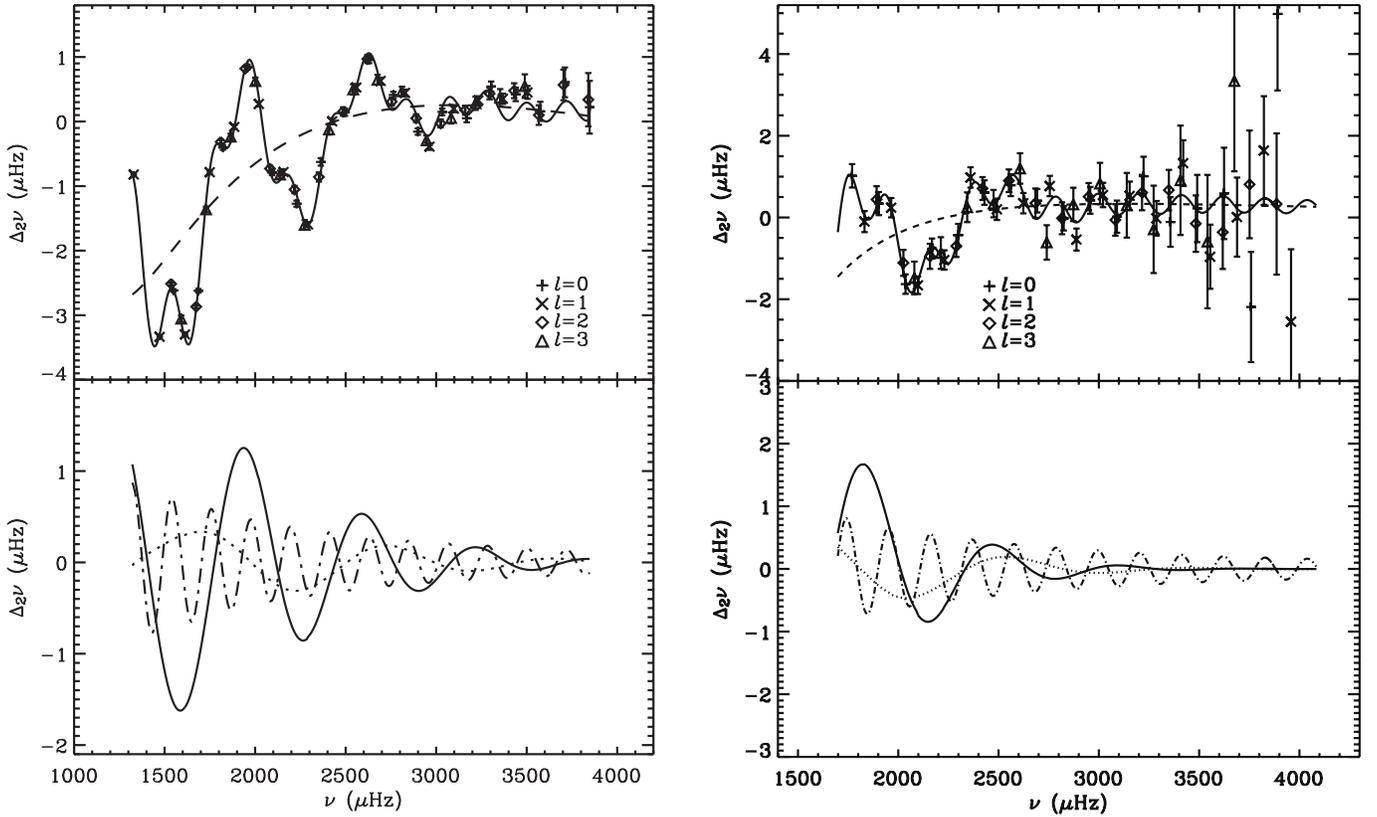}
     \end{center}
     \caption{The symbols in the {\bf upper panels} denote second differences
          $\Delta_{2i}\nu:=\nu_{n-1,l}-2\nu_{n,l}+\nu_{n+1,l}$ for low-degree 
          modes obtained from the BiSON ({\bf left panel}; Basu et al. 2007) 
          and from simulated data for a 1\,M$_\odot$ model of age 5.54 Gy 
          ({\bf right panel}); the simulation was based on two periods 
          of 4-months of observation, separated by 1 year, with the SONG 
          network (see Grundahl et al. 2007). The solid curves are fits to 
          $\Delta_{2i}\nu$ based on the analysis by Houdek \& Gough (2007). The 
          dashed curves are the smooth contributions, including a third-order 
          polynomial in $\nu^{-1}_i$ to represent the upper-glitch contribution
          from near-surface effects. The {\bf lower panels} display the 
          remaining individual contributions from the acoustic glitches to 
          $\Delta_{2i}\nu$: the dotted and solid curves are the contributions 
          from the first and second stages of helium ionization, and the 
          dot-dashed curve is the contribution from the acoustic glitch at the 
          base of the convective envelope.}
     \label{fig:two_columns}
   \end{figure*}

\begin{eqnarray}
\delta_{{\rm u}}\nu_i
&\simeq&A+B\nu_i\cr
&\!\!+\!\!&{\left[\frac{1}{2}a_0\nu_i^2+a_1\nu_i(\ln\nu_i\!-\!1)-a_2\ln\nu_i+\frac{1}{2}a_3\nu_i\right]h^{-2}}\cr
&\equiv&A+B\nu_i+F_{\rm u}\,,
\label{e:delnu_surf}
\end{eqnarray}
where $ h=(\nu_{n+1,l}-\nu_{n-1,l})/2$. The somewhat arbitrarily chosen power 
series in $\nu^{-1}_i$ for the upper-glitch component leads to the two 
undetermined constants of summation $A$ and $B$, which \citet{Houdek09} chose 
by minimizing
\begin{equation}
E\equiv||{\delta_{\rm u}\nu_i}||_2=\sum_n(A+B\nu_i+F_{\rm u})^2\,.
\end{equation}
In Fig.\,1 is displayed the sum of all acoustic glitches, i.e. 
Eq.(\ref{e:all_glitches}), for low-degree solar frequencies observed by 
BiSON \citep{Basu07}. Displayed in the top panels of Fig.\,2 are the 
second differences of $\delta\nu_i$ (solid curve) and of the upper-glitch 
component $\delta_{\rm u}\nu_i$ (dashed curve) for the Sun and for simulated
data of a solar-like star observed with the Danish SONG instrument. The 
corresponding individual frequency contributions $\delta_\gamma\nu_i$ (dotted 
and solid curves for the two stages of helium ionization) and 
$\delta_{\rm c}\nu_i$ (dot-dashed curve) are illustrated in the lower panels 
of Fig.\,2.

The amplitudes of the oscillatory contributions from the helium ionization 
(solid and dotted curves in the lower panels of Fig.\,2) provide an independent 
measure of the helium abundance of the star. Moreover, with the help of the
smoothed (glitch-free) eigenfrequencies $\nu_{{\rm s}i}$ the core-sensitive
coefficients of the asymptotic representation of solar-like oscillations
\citep{Tassoul80, Gough86} can be evaluated more accurately. This 
information stabilizes the age calibration of the Sun and solar-like stars 
using only low-degree acoustic modes \citep{Houdek08, Houdek09}.

\begin{acknowledgements}
Support by the Austrian Science Fund (FWF project P21205) is gratefully 
acknowledged. Participation at the Ponte de Lima workshop was supported by 
the European Helio- and Asteroseismology Network (HELAS), a major 
international collaboration funded by the European Commission's Sixth 
Framework Programme.
\end{acknowledgements}

\bibliographystyle{aa}
\bibliography{biblio}

\begin{thebibliography}{}

\bibitem[Basu et al.(2007)]{Basu07}
Basu S., Chaplin, W.~J., Elsworth, Y., New, A.~M., Serenelli, G., 
Verner, G.~A. 2007, ApJ, 655, 660

\bibitem[Gough(1986)]{Gough86}
Gough D.O. 1986, in: Proc. Hydrodynamic and magnetohydrodynamic
problems in the Sun and stars, ed. Y. Osaki, University of 
Tokyo, Tokyo, p.\,117 

\bibitem[Grundahl et al.(2007)]{Grundahl07}
Grundahl F., Kjeldsen H., Christensen-Dalsgaard J., Arentoft T., 
Frandsen S. 2007, CoAst 150, 300

\bibitem[Houdek \& Gough(2007)]{Houdek07}
Houdek G., Gough D.O. 2007, \mnras\ 375, 861.

\bibitem[Houdek \& Gough(2008)]{Houdek08}
Houdek G., Gough D.O. 2008, In {The Art of Modelling
Stars in the 21st Century}, eds Deng, L., Chan, K.~L., Chiosi C.,
IAU Symp., Vol.\,{252}, CUP, Cambridge , p.~149.

\bibitem[Houdek \& Gough(2009)]{Houdek09}
Houdek G., Gough D.O. 2009, CoAst 159, 27

\bibitem[Tassoul(1980)]{Tassoul80}
Tassoul M. 1980, ApJS 43, 469

\end{thebibliography}

\end{document}